\documentclass[conference]{IEEEtran}
\IEEEoverridecommandlockouts
\usepackage{cite}
\usepackage{amsmath,amssymb,amsfonts}
\usepackage{algorithmic}
\usepackage{graphicx}
\usepackage{subfigure}
\usepackage{etoolbox}
\makeatletter
\patchcmd{\@makecaption}
  {\scshape}
  {}  {}  {}
\makeatother
\usepackage{multirow}
\usepackage[table,xcdraw]{xcolor}
\usepackage{textcomp}
\usepackage{xcolor}
\def\BibTeX{{\rm B\kern-.05em{\sc i\kern-.025em b}\kern-.08em
    T\kern-.1667em\lower.7ex\hbox{E}\kern-.125emX}}
\begin{document}

\title{CNN-based Prediction of Network Robustness With Missing Edges\\
\thanks{This research was supported by the National Natural Science Foundation of China (No. 62002249) and the Foundation of Key Laboratory of System Control and Information Processing, Ministry of Education, P. R. China (No. Scip202103).}}
\author{\IEEEauthorblockN{Chengpei Wu\textsuperscript{1},
Yang Lou\textsuperscript{1,2},
Ruizi Wu\textsuperscript{1},
Wenwen Liu\textsuperscript{1},
and Junli Li\textsuperscript{1}}
\IEEEauthorblockA{1. \textit{College of Computer Science, Sichuan Normal University, Chengdu 610066, China}\\
2. \textit{Department of Computing and Decision Sciences, Lingnan University, Hong Kong, China}\\
e-mails: \textit{felix.lou@ieee.org}, \textit{lijunli@sicnu.edu.cn}
}}

\maketitle
\begin{abstract} 
Connectivity and controllability of a complex network are two important issues that guarantee a networked system to function. Robustness of connectivity and controllability guarantees the system to function properly and stably under various malicious attacks. Evaluating network robustness using attack simulations is time consuming, while the convolutional neural network (CNN)-based prediction approach provides a cost-efficient method to approximate the network robustness.
In this paper, we investigate the performance of CNN-based approaches for connectivity and controllability robustness prediction, when partial network information is missing, namely the adjacency matrix is incomplete.  
Extensive experimental studies are carried out. A threshold is explored that if a total amount of more than 7.29\% information is lost, the performance of CNN-based prediction will be significantly degenerated for all cases in the experiments. 
Two scenarios of missing edge representations are compared, 1) a missing edge is marked `no edge' in the input for prediction, and 2) a missing edge is denoted using a special marker of `unknown'. Experimental results reveal that the first representation is misleading to the CNN-based predictors.
\end{abstract}

\begin{IEEEkeywords}
Complex network, convolutional neural network, robustness, prediction, missing edge.
\end{IEEEkeywords}

\section{Introduction}
Many nature and engineering systems can be modelled as complex networks and studied using various complex network analysis tool. The study of complex networks covers multiple disciplines, including not only mathematics, physics, computer science, but also social sciences, network sciences and biological sciences\cite{Newman2010N,Chen2014Book,Barabasi2016NS,Chen2019Book}.

Connectivity is fundamental to a complex network, which ensures that the system can be considered as a whole. Connectivity is guaranteed by a sufficient number of edges that connect nodes properly. Network connectivity is important to real-world systems such as power grid\cite{Cuadra2015ENG} and transportation network\cite{Wandelt2021RESS}.
Controllability refers to the ability of a networked system can be steered from any initial state to any target state under an admissible control input, within a finite duration of time. 

Recently, researches on network robustness have attracted increasing attention\cite{Pu2012PA,Sun2019ICSRS,Lou2021CNSNS,Sun2021TNSM}, since malicious attacks and random failures become inevitable. Network robustness reflects the ability of a complex network to maintain or retain its structure and functions. In this paper, specifically, network robustness refers to connectivity robustness and controllability robustness.

Connectivity robustness and controllability robustness are measured by recording the changes of network connectivity and controllability under a series of node- or edge-removal attacks. The percolation theory\cite{2021Percolation} implies that the largest connected component (LCC) plays an important role in maintaining the network structure, and a widely-used measure is proposed based on the changes of the proportion of LCC for connectivity robustness measure\cite{Schneider2011PNAS}.
As for controllability robustness\cite{Chen2019TCASII}, the changes of proportion of driver nodes are recorded as the measure. Conventionally, measuring both connectivity robustness and controllability robustness require time-consuming simulation, while the application of easy-to-access indicators such as assortativity\cite{Newman2003PRE} and spectral measures\cite{Perra2008PRE} have limited scopes of applications, and therefore time-consuming attack simulations remain as the main approach today.

% deep learning
Deep learning has performed great potential in data mining. As a data-driven method, deep neural networks are capable to learn comprehensive data features without human intervention. Deep learning has been also widely applied in processing complex network data. For example, graph attention network (GAT)\cite{Petar2018graph} is employed to find out the hidden key nodes with maximum influences to the network\cite{grassia2021machine}. Deep reinforcement learning is also used to find out a set of key players in real-world networks\cite{Fan2020NMI}.

Convolutional neural network (CNN) has shown powerful capability in image processing\cite{Schmidhuber2015NN}, which also provides a useful tool for robustness prediction. An adjacency matrix of network can be converted into an image with one channel, and many existing synthetic network generation models provide a sufficiently large number of synthetic images for data mining. Therefore, complex network data can be processed by CNNs using an image processing manner\cite{Lou2020TCYB,Lou2021TNNLS,Lou2021TNSE}. Although the CNN-based prediction approach performs well on network robustness prediction, as well as network classification\cite{Lou2021TNNLS}. it requires a full-knowledge of complex network. A non-trivial change in network size may degenerate the performance.  In contrast, it is common that the data of real-world networks are incomplete, where missing nodes and/or edges are common forms of information loss.

In this paper, we investigate the robustness prediction performance of CNN-based approach, where the network structure information is incomplete. Specifically, the scenario of missing edges is taken into account in the CNN-based robustness prediction.

The rest of the paper is organized as follows: Section \ref{sec:pre} introduces some preliminaries, including basic concepts and definitions. Section \ref{sec:cnn} elaborates in detail the CNN-based robustness prediction with missing information. Section \ref{sec:exp} demonstrates simulation results with analysis and discussions. Finally, Section \ref{sec:end} concludes the investigation.

\section{Preliminaries}\label{sec:pre}

In this paper, two robustness measures of directed networks are predicted and investigated, namely the \textit{controllability} robustness and \textit{connectivity} robustness. The former reflects how well a networked system can maintain or regain its controllable state, while the later reflects how well it can maintain its connectedness, both against destructive attacks.  Only node-removal attacks are considered in this paper, while edge-removal attacks can be studied in the same way.

\subsection{Connectivity Robustness}

Connectivity is fundamentally important for networks to function. For a directed network, it is \textit{weakly connected}, if it remains to be \textit{connected} after all the directions are removed, where connected means that for each pair of nodes there is an undirected path between them. 

Network connectivity robustness under node-removal attacks is widely recorded as a sequence of fractions of nodes in the largest connected component (LCC)\cite{Schneider2011PNAS}, or a normalized LCC (NLCC) curve. Each NLCC value is calculated as follows:
\begin{equation}\label{eq:nlc}
    s(i)=\frac{c(i)}{N-i}\,,~~i=0,1,\ldots,N-1\,,
\end{equation}
where $c(i)$ is the number of nodes in LCC after a total number of $i$ nodes have been removed from the network; $s(i)$ is the NLCC value after a total number of $i$ nodes removed; $N$ is the number of nodes in the original network before being attacked. When these values are plotted, a curve is obtained, called the \textit{connectivity curve}.

Connectivity guarantees the fundamental functionalities of a complex network, including such as controllability\cite{Xiang2019CSM}, synchronizability\cite{Shi2013CSM}, and the abilities of communication and transmission, etc.

\subsection{Controllability Robustness}

For a linear time-invariant networked system $\dot{{\bf x}}=A{\bf x}+B{\bf u}$, where $A$ and $B$ are constant matrices of compatible dimensions, $\bf x$ and $\bf u$ are the state vector and control input, respectively. The system is \textit{state controllable} if and only if the controllability matrix $[B\ AB\ A^2B\ \cdots A^{N-1}B]$ has a full row-rank, where $N$ is the dimension of $A$, also the size of the network. It is shown\cite{Liu2011N} that, for a directed network, identifying the set of the minimum number of external driver nodes $N_D$ can be converted to searching for a maximum matching of the network, namely $N_D$ can be calculated by $N_D=\text{max}\{1, N-|E^*|\}$, where $|E^*|$ is the number of edges in the maximum matching $E^*$\cite{Liu2011N}.

The measure of controllability robustness is calculated by
\begin{equation}\label{eq:ndi}
	n_D(i)=\frac{N_D(i)}{N-i}\,,\ \ i=0,1,\ldots,N-1,
\end{equation}
where $N_D(i)$ is the number of driver nodes needed to retain the network controllability after a total of $i$ nodes have been removed, and $N$ is the original network size. When these values are plotted, a curve is obtained, called the \textit{controllability curve}.

\section{Convolutional Neural Networks for Robustness Prediction}\label{sec:cnn}
\subsection{CNN-based Predictor}
The CNN-based robustness predictor proposed in\cite{Lou2021TNSE} is employed to predict connectivity robustness, the predictor proposed in \cite{Lou2020TCYB} is employed to predict the controllability robustness. In this paper, these two predictors are given a unified name CNN-based robustness predictor (CNN-RP).

\begin{figure}[htbp]
	\centering
	\includegraphics[width=1\linewidth]{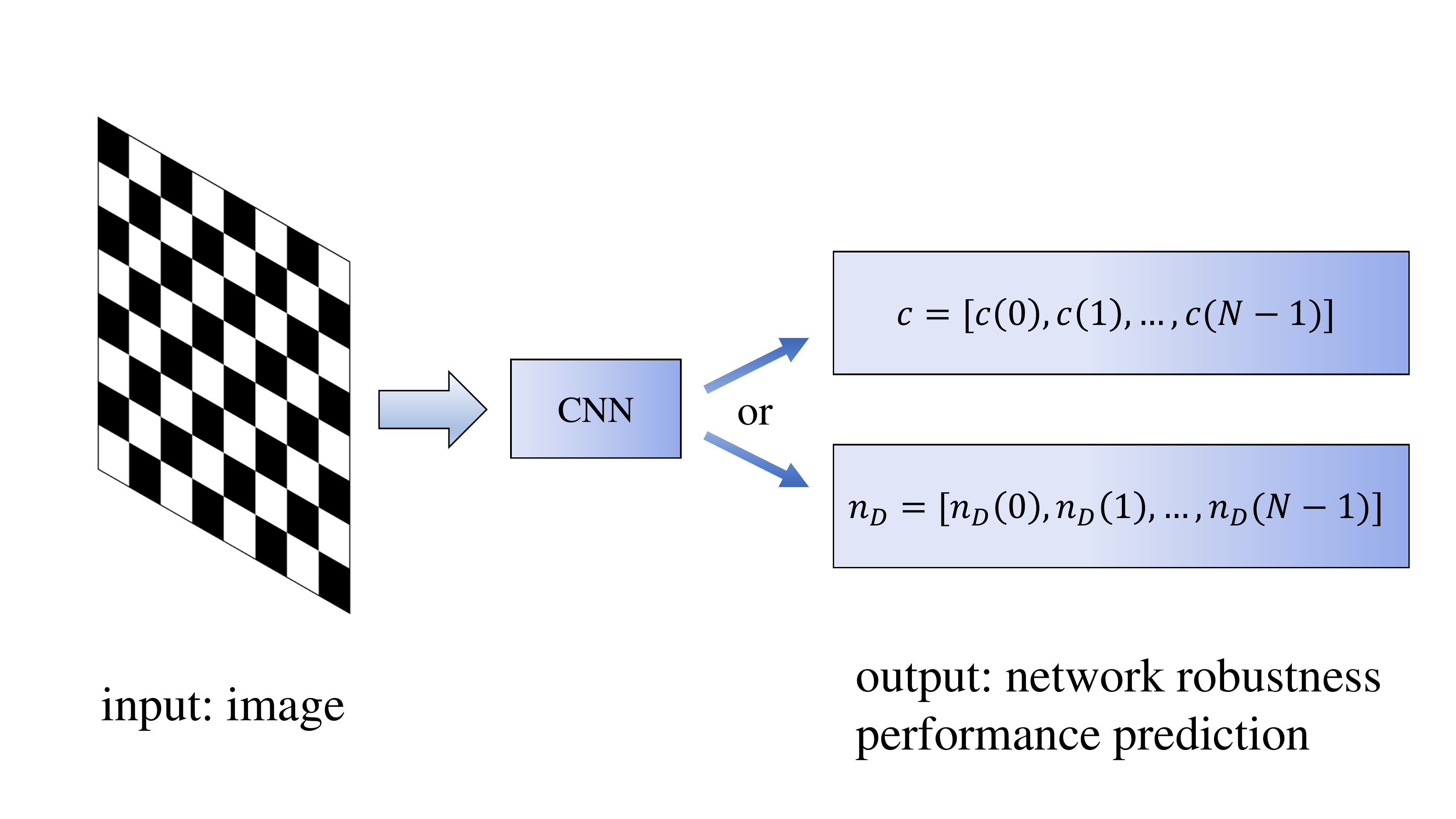}
	\caption{Framework of CNN-RP: The input is an adjacency matrix converted image; the output is a predicted LCC curve $c$ or a controllability curve $n_D$.}
	\label{fig:cnn}
\end{figure}

As shown in Fig. \ref{fig:cnn}, the gray-scale image converted from adjacency matrix is used as the input, and the output is the predicted robustness performance. Only directed unweighted networks are investigated in this paper, thus each converted image contains only black and white pixels. A black pixel represents a `0' in adjacency matrix, while a white pixel represents a `1', which represents the existence of an edge between the corresponding pair of nodes.

\begin{figure}[htbp]
	\centering \includegraphics[width=\linewidth]{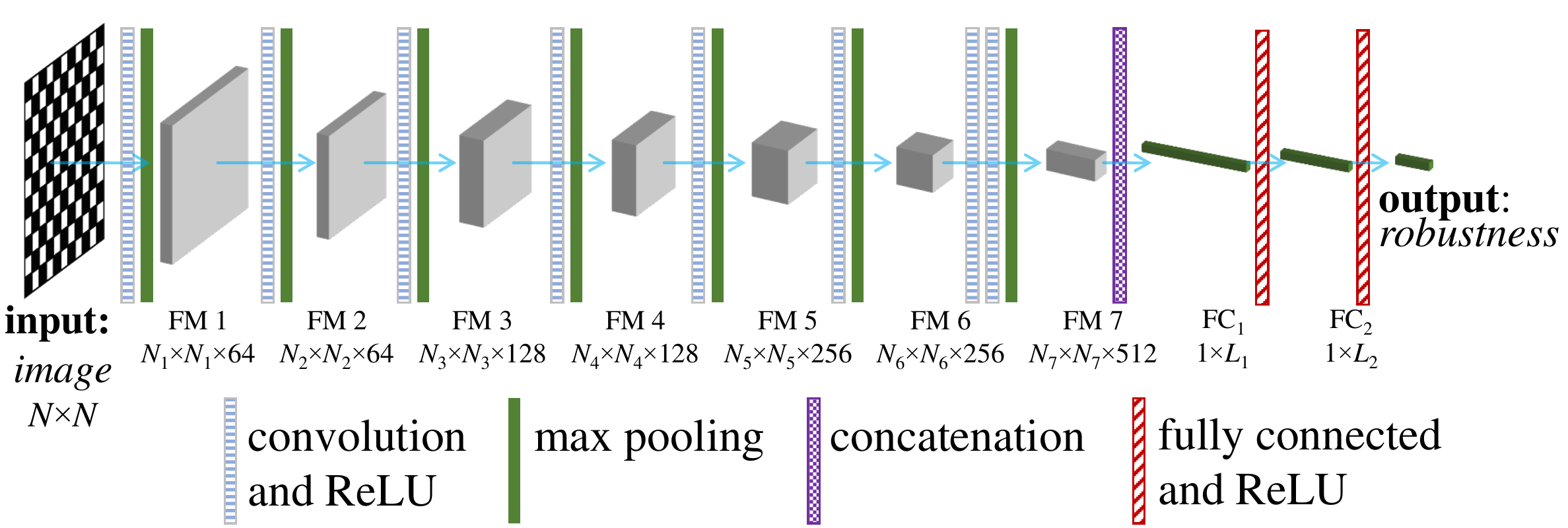}
	\caption{CNN structure of  CNN-RP. The input is gray-scale image; the output is robustness performance. }
	\label{fig:cnn_struct}
\end{figure}

The CNN structure of CNN-RP is shown in Fig. \ref{fig:cnn_struct}. The detailed configurations and parameters can be referred to \cite{Lou2020TCYB,Lou2021TNSE}.
The mean-squared error between the predicted connectivity or controllability curve $\hat{v}$ and true curve $v$ is used as the loss function:
\begin{equation}\label{eq:lf}
	\mathcal{L} = \frac{1}{N} \sum_{i=0}^{N-1}||\hat{v}(i)-v(i)||\,,
\end{equation}
where $\hat{v}(i)$ represents the predicted connectivity (see Equation (\ref{eq:nlc})) or controllability (see Equation (\ref{eq:ndi})) curve, while $v(i)$ represents the corresponding true curve; $||\cdot||$ represents the Euclidean norm. The training process of CNN-RP aims at adjusting the internal parameters, with the objective of minimizing $\mathcal{L}$.

\subsection{Missing Information}

Given an $N$-node network, an $N\times N$ gray-scale image can be generated from its adjacency matrix. This is implemented by converting and scaling the element values of a adjacency matrix to gray-scale pixel values.  Missing information is implemented by adding an $S\times S$ square mask onto the gray-scale image, where the edge information within the mask range will be unseen.

The location $P$ of a mask is defined as the upper-left corner. For each mask, its location is uniformly-randomly set within a gray-scale image, as follows:
\begin{equation}\label{eq:position}
    P\in\{(i,j);~i,j=\text{rand}(1,N-S)\} 
\end{equation}
where $\text{rand}(a,b)$ represents a random integer in the range $[a,b]$.

Two strategies of missing edge notations are implemented and compared. The first strategy is to assign all `0's to the mask. Thus, all the existing edges (`1's) within the mask-covered area will be assigned `0's. As a result, this mask actually gives a misleading information about the existence of edges in the covered area. This mask is called a \textit{null mask}.
A null mask may mislead observers (either people or CNN-based predictor) by covering all edges within the mask-covered area.

The second mask strategy is to assign specific values that is different from `0' or `1'. Such a mask is named a \textit{confusion mask}. In the implementation, a value of 0.5 is assigned for each pixel in the masked square area. A confusion mask may be less misleading than a null mask, since the non-zero non-one value can be easier recognized by observers. However, the edge information within the mask-covered area remains unseen.

\begin{figure}[htbp]
	\centering
	\includegraphics[width=\linewidth]{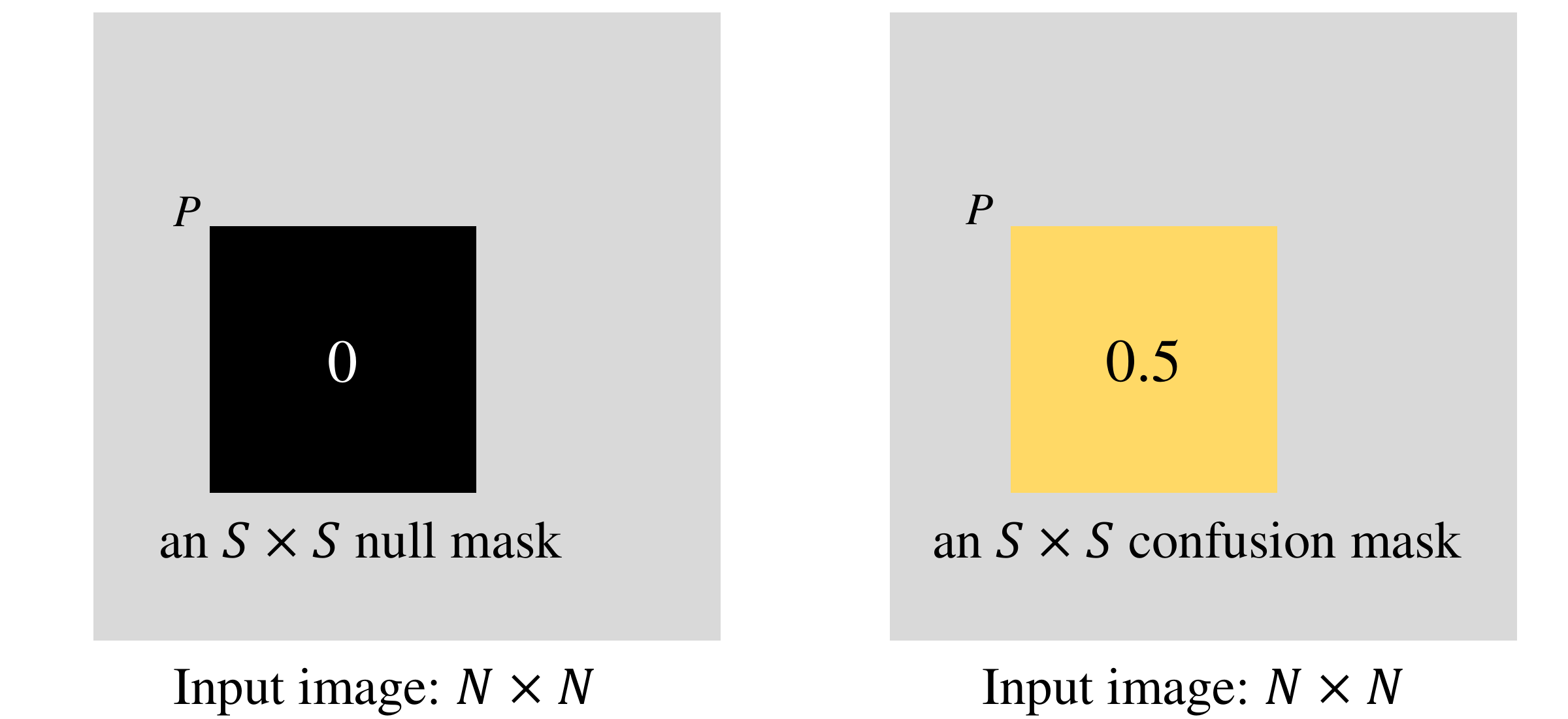}
	\caption{An example of null mask and confusion mask.}
	\label{fig:masks}
\end{figure}

\section{Experimental Studies}\label{sec:exp}
\subsection{General Experimental Settings}
Four synthetic directed network models are employed for the experiments, including Erd{\"{o}}s--R{\'e}nyi (ER)\cite{Erdos1964RG}, \textit{q}-snapback (QS)\cite{Lou2018TCASI,Lou2019R}, Newman--Watts small-world (SW)\cite{Newman1999PLA}, and the generic scale-free (SF)\cite{Goh2001PRL} networks. The number of nodes is set to $N=1000$. The average degree $\langle k\rangle$ is set to 4, 7 and 9, respectively. 
Both connectivity robustness and controllability robustness are studied, under three typical node-removal attack strategies including random attacks (RA), targeted betweenness-based attacks (TB), and targeted degree-based attacks (TD).

Two experiments are implemented. In Experiment I, null masks are used as the missing information for complex networks. This experiment explores the prediction performance of CNN-RP when the size of null mask changes. In Experiment II, confusion masks are compared with null masks, which investigates whether the CNN-based approach can learn to recognize and deal with a certain region where there are possible missing edges.

\subsection{Experiment I}
In Experiment I, the number of training samples is $6000=4\times3\times500$, namely, 4 network topologies, 3 different degree settings, and 500 random network instances for each network configuration.  Note that there is no information missing in the training data. The number of testing networks is $1200=4\times3\times100$, representing 4 network topologies, 3 different degree settings, and 100 random instances for each configuration.  For each testing network, a null mask with a fixed size is covered onto the converted image at a random location. Different mask sizes are also tested.

\begin{figure*}[htbp]
	\centering
	\includegraphics[width=.8\linewidth]{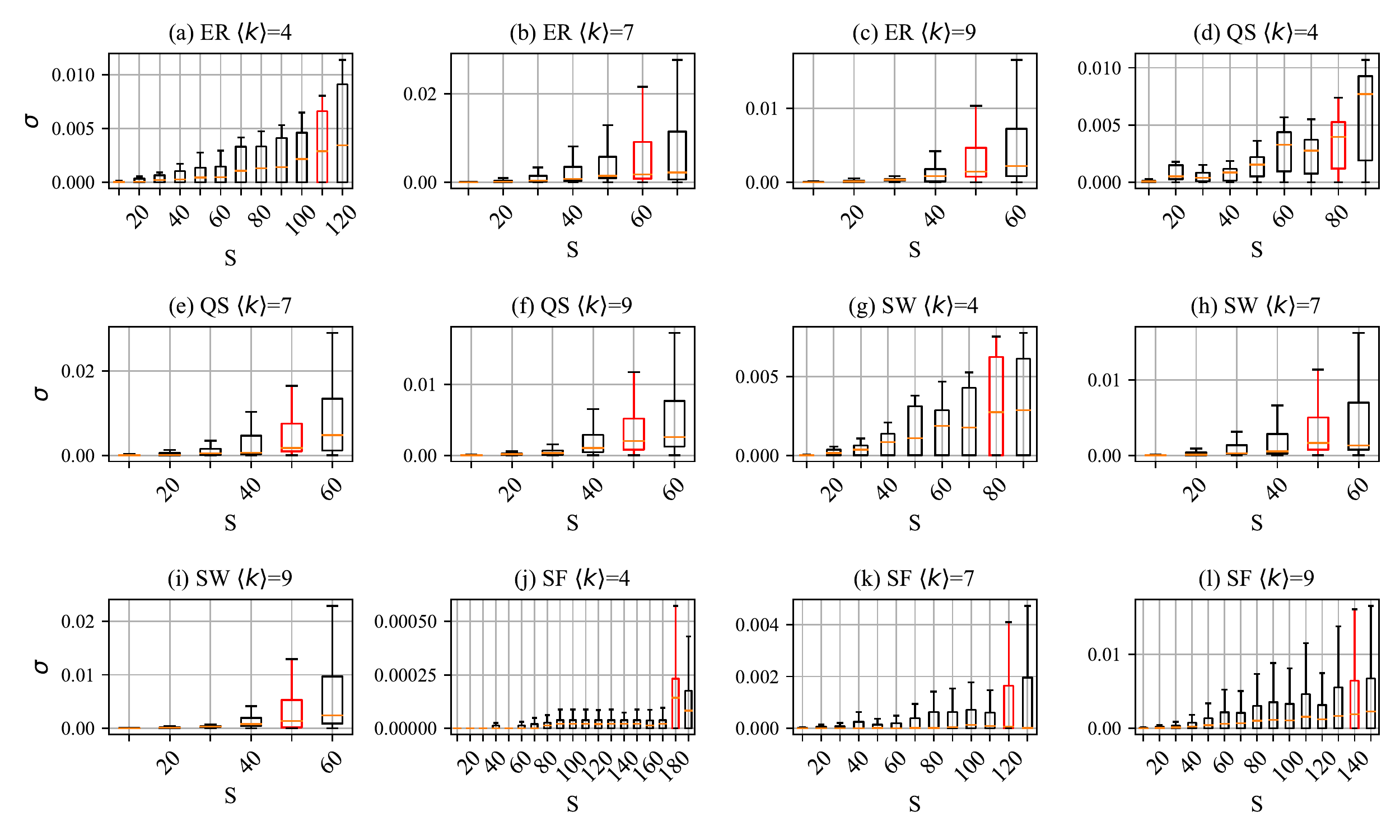}
	\caption{Box plot of CNN-RP connectivity robustness prediction for ER, QS, SW and SF networks under TB attacks. $\sigma$ represents the average error coursed by edge information loss. $S$ represents the size of null mask.}
	\label{fig:l_tb}
\end{figure*}

\begin{figure*}[htbp]
	\centering
	\includegraphics[width=.8\linewidth]{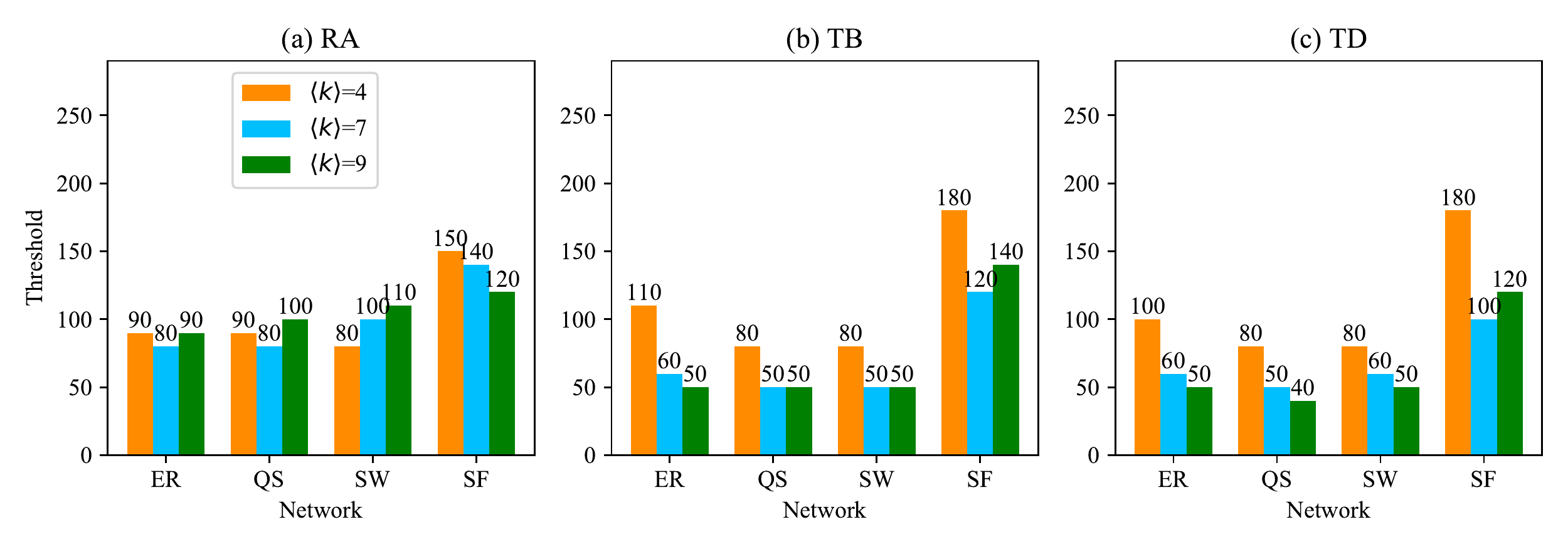}
	\caption{Threshold of null mask size, for CNN-RP connectivity robustness prediction.}
	\label{fig:threshold_connectivity}
\end{figure*}

\begin{figure*}[htbp]
	\centering
	\includegraphics[width=.8\linewidth]{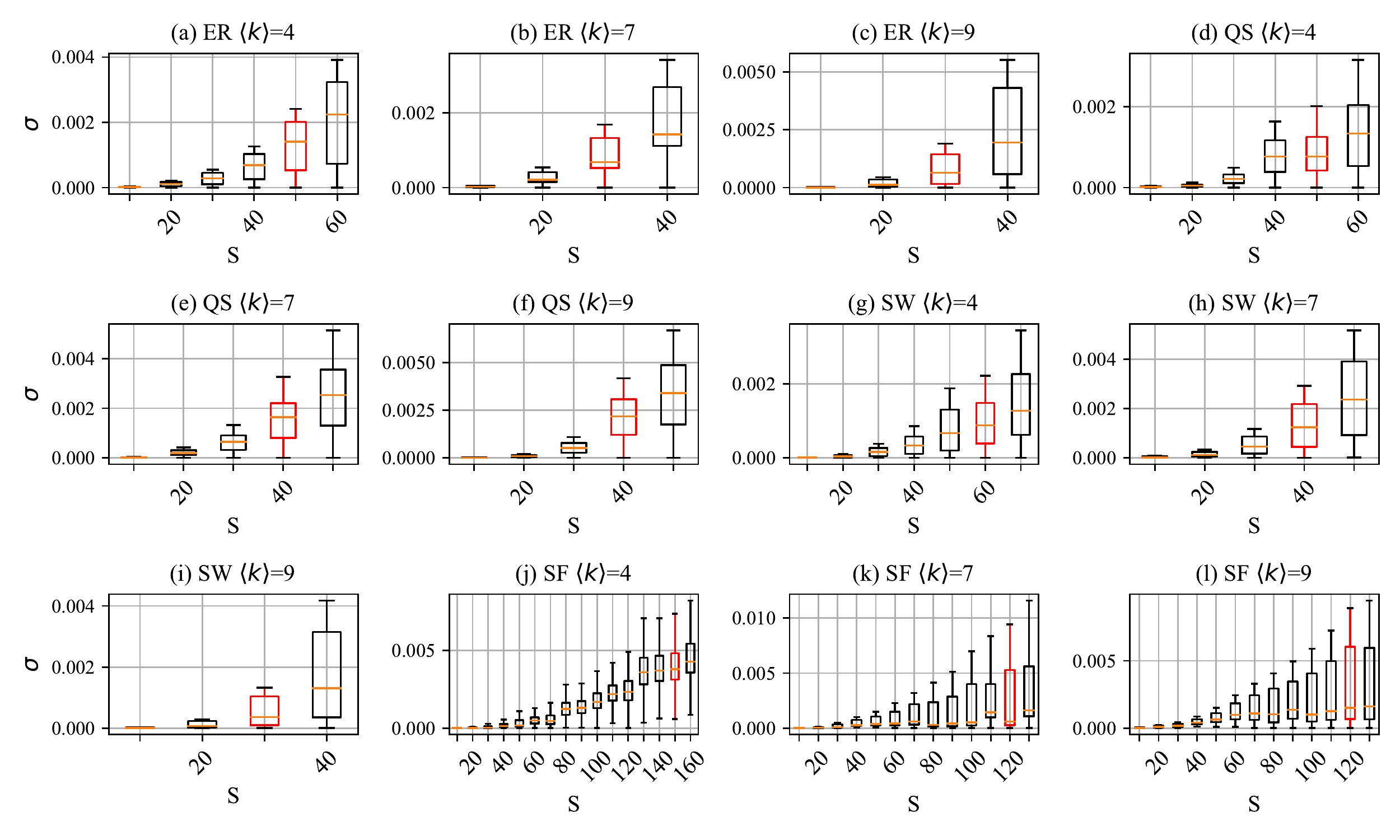}
	\caption{Box plot of CNN-RP controllability robustness prediction for ER, QS, SW and SF networks under TB attacks. $\sigma$ represents the average error coursed by edge information loss. $S$ represents the size of null mask.}
	\label{fig:c_tb}
\end{figure*}

\begin{figure*}[htbp]
	\centering
	\includegraphics[width=.8\linewidth]{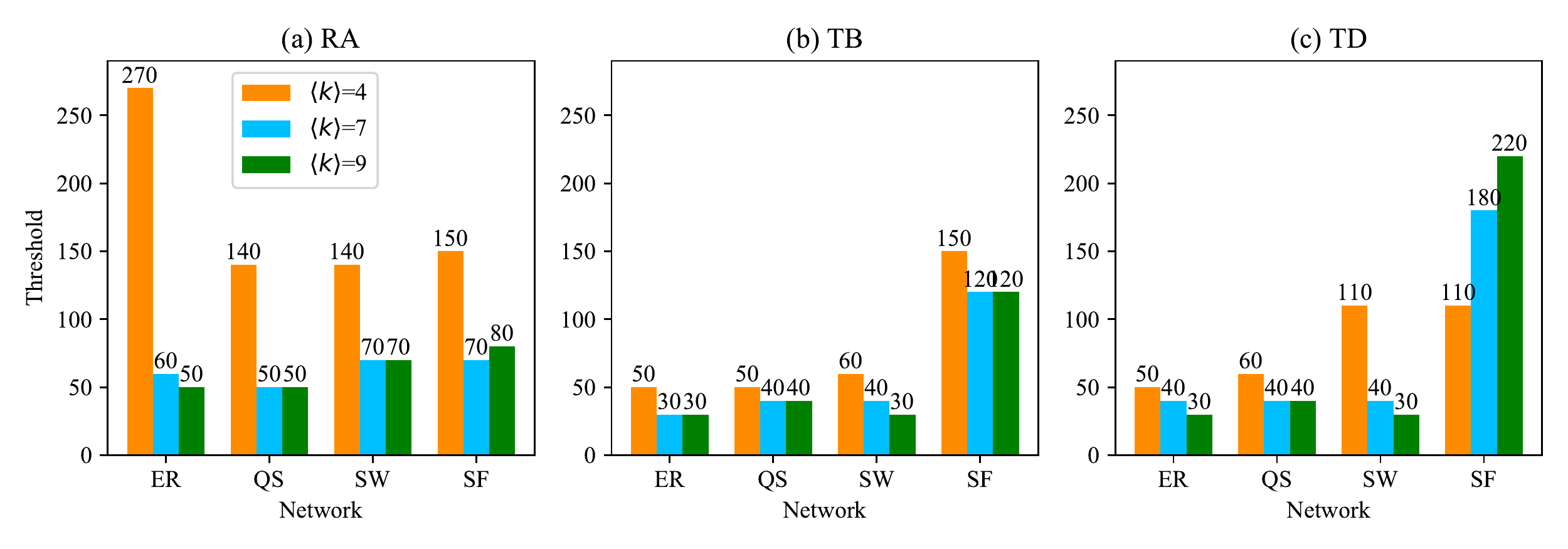}
	\caption{Threshold of null mask size, for CNN-RP controllability robustness prediction.}
	\label{fig:threshold_controllability}
\end{figure*}

Fig. \ref{fig:l_tb} shows the box plot of CNN-RP connectivity robustness prediction for ER, QS, SW and SF networks under TB attacks. $\sigma$ represents the average error coursed by edge information loss. When there is a statistic significance detected, the corresponding box is highlighted in red. For example, in Fig. \ref{fig:l_tb} (a), for ER networks with an average degree $\langle k\rangle=4$, as the size of null mask $S$ increases, the average error of connectivity robustness prediction increases, while the increase of error is insignificant until $S$ reaches 110, when a significant difference is detected. This means that a null mask of $S\geq110$ will significantly degenerate the prediction performance of CNN-RP. The Mann-Whitney U-test (with $\alpha=0.05$) is employed for the hypothesis test.

The thresholds of null mask size for different networks under RA, TB and TD attack strategies are shown in Fig. \ref{fig:threshold_connectivity}. Specifically, the null mask size starts from 10, with incremental of 10. Hypothesis test is performed for each mask size to check whether the current size has significantly degenerate the prediction performance. 
A threshold is determined when the prediction error is significantly increased.

A greater threshold value means that CNN-RP has greater tolerance of information loss in that case; while a lower threshold means that CNN-RP is more sensitive to the information loss.  As shown in Fig. \ref{fig:threshold_connectivity}, for ER networks with $\langle k\rangle=4$ under RA attacks, CNN-RP can resist a mask size of $S=110$. The ratio of pixel loss is $1.21\%$.

Fig. \ref{fig:c_tb} shows the box plot of CNN-RP controllability robustness prediction for ER, QS, SW and SF networks under TB attacks. The corresponding thresholds of null mask size is shown in Fig. \ref{fig:threshold_controllability}.

As can be seen from Fig. \ref{fig:threshold_connectivity}, for connectivity robustness prediction, among all the network configurations and different attack strategies, the minimum threshold of mask size is 40 (QS with $\langle k\rangle=9$ under TD attacks, as shown in Fig. \ref{fig:threshold_connectivity} (c)). The maximum threshold is 180 (SF with $\langle k\rangle=4$ under TB and TD attacks, as shown in Figs. \ref{fig:threshold_connectivity} (b) and (c), respectively).  From Fig. \ref{fig:threshold_controllability}, for controllability robustness prediction, the minimum threshold of mask size is 30 (ER with $\langle k\rangle=7$ under TB, ER with $\langle k\rangle=9$ under TB and TD, and SW with $\langle k\rangle=9$ under TB and TD attacks, as shown in Figs. \ref{fig:threshold_connectivity} (b) and (c)). The maximum threshold is 270 (ER with $\langle k\rangle=4$ under RA attacks, as shown in Fig. \ref{fig:threshold_connectivity} (a)).

Overall, for both connectivity and controllability robustness prediction, the ratios of pixel loss for the minimum and maximum thresholds are 0.09\% (when $S=30$) and 7.29\% (when $S=270$), respectively.

\subsection{Experiment II}

In Experiment II, both the original complete networks and mask-covered networks are employed in the training set.   Similar to Experiment I, 4 network topologies with 3 average degrees are set. For each network configuration, 4 random network instances are generated as the original network. Thus, there are 48 network instances (without masks) for each network configuration. 
Both null and confusion masks with different sizes will be randomly imposed over the adjacency matrices, and 100 random masked instances are generated. Thus, there are 4800 masked network instances. thus, the total numbers of training networks are 4848, 4848, and 1968 for three different mask sizes 54, 216, and 360, respectively. 
CNN-RP models are trained independently for different mask sizes.

The objective of Experiment II is to compare the influences of null masks and confusion masks to the prediction performance of CNN-RP.  The performance difference is compared by using the mean absolute error (MAE) of the true curve and the predicted curve. A lower MAE value means better prediction performance.

\begin{table*}[htbp]
\centering
\caption{Difference of MAE values of connectivity prediction between the null-mask and confusion-mask information loss, when $\langle k\rangle=4$.}
\begin{tabular}{|c|ccc|ccc|ccc|ccc|}
\hline & \multicolumn{3}{c|}{ER} & \multicolumn{3}{c|}{QS} & \multicolumn{3}{c|}{SW} & \multicolumn{3}{c|}{SF} \\ \cline{2-13} 
\multirow{-2}{*}{} & \multicolumn{1}{c|}{54} & \multicolumn{1}{c|}{216} & 360 & \multicolumn{1}{c|}{54} & \multicolumn{1}{c|}{216} & 360 &\multicolumn{1}{c|}{54} & \multicolumn{1}{c|}{216} & 360 & \multicolumn{1}{c|}{54} & \multicolumn{1}{c|}{216} & 360 \\ \hline
RA & \multicolumn{1}{c|}{\cellcolor[HTML]{C9C9C9}--1.34e--4} & \multicolumn{1}{c|}{2.76e--3} & 1.27e--4 & \multicolumn{1}{c|}{9.68e--5} & \multicolumn{1}{c|}{\cellcolor[HTML]{C9C9C9}--9.40e--5} & 1.44e--3 & \multicolumn{1}{c|}{3.54e--4} & \multicolumn{1}{c|}{7.00e--3} & 1.50e--3 & \multicolumn{1}{c|}{\cellcolor[HTML]{C9C9C9}--2.73e--4} & \multicolumn{1}{c|}{1.53e--4} & 9.38e--4 \\ \hline
TB & \multicolumn{1}{c|}{\cellcolor[HTML]{C9C9C9}--3.15e--5} & \multicolumn{1}{c|}{2.19e--3} & \cellcolor[HTML]{C9C9C9}--2.22e--2 & \multicolumn{1}{c|}{8.70e--6} & \multicolumn{1}{c|}{\cellcolor[HTML]{C9C9C9}--8.61e--4} & \cellcolor[HTML]{C9C9C9}--2.23e--2 & \multicolumn{1}{c|}{2.86e--4} & \multicolumn{1}{c|}{1.74e--3} & \cellcolor[HTML]{C9C9C9}--1.70e--2 & \multicolumn{1}{c|}{7.53e--5} & \multicolumn{1}{c|}{\cellcolor[HTML]{C9C9C9}--1.87e--3} & \cellcolor[HTML]{C9C9C9}--1.52e--2 \\ \hline
TD & \multicolumn{1}{c|}{\cellcolor[HTML]{C9C9C9}--4.96e--5} & \multicolumn{1}{c|}{\cellcolor[HTML]{C9C9C9}--3.93e--3} & 1.53e--2 & \multicolumn{1}{c|}{\cellcolor[HTML]{C9C9C9}--8.39e--4} & \multicolumn{1}{c|}{\cellcolor[HTML]{C9C9C9}--4.13e--3} & 1.86e--2 & \multicolumn{1}{c|}{5.68e--4} & \multicolumn{1}{c|}{\cellcolor[HTML]{C9C9C9}--4.27e--3} &
  1.54e--2 & \multicolumn{1}{c|}{2.93e--5} & \multicolumn{1}{c|}{\cellcolor[HTML]{C9C9C9}--7.77e--4} & 1.06e--3 \\ \hline
\end{tabular}
\label{tab:difference_connectivity4}
\end{table*}

\begin{table*}[htbp]
\centering
\caption{Difference of MAE values of controllability prediction between the null-mask and confusion-mask information loss, when $\langle k\rangle=4$.}
\begin{tabular}{|c|ccc|ccc|ccc|ccc|}
\hline
 &
  \multicolumn{3}{c|}{ER} &
  \multicolumn{3}{c|}{QS} &
  \multicolumn{3}{c|}{SW} &
  \multicolumn{3}{c|}{SF} \\ \cline{2-13} 
\multirow{-2}{*}{} &
  \multicolumn{1}{c|}{54} &
  \multicolumn{1}{c|}{216} &
  360 &
  \multicolumn{1}{c|}{54} &
  \multicolumn{1}{c|}{216} &
  360 &
  \multicolumn{1}{c|}{54} &
  \multicolumn{1}{c|}{216} &
  360 &
  \multicolumn{1}{c|}{54} &
  \multicolumn{1}{c|}{216} &
  360 \\ \hline
RA &
  \multicolumn{1}{c|}{\cellcolor[HTML]{C9C9C9}--9.99e--5} &
  \multicolumn{1}{c|}{2.14e--2} &
  1.66e--2 &
  \multicolumn{1}{c|}{\cellcolor[HTML]{C9C9C9}--6.84e--4} &
  \multicolumn{1}{c|}{2.04e--2} &
  1.69e--2 &
  \multicolumn{1}{c|}{\cellcolor[HTML]{C9C9C9}--8.21e--4} &
  \multicolumn{1}{c|}{2.13e--2} &
  1.47e--2 &
  \multicolumn{1}{c|}{5.45e--4} &
  \multicolumn{1}{c|}{1.23e--2} &
  \cellcolor[HTML]{C9C9C9}--7.44e--3 \\ \hline
TB &
  \multicolumn{1}{c|}{\cellcolor[HTML]{C9C9C9}--4.76e--4} &
  \multicolumn{1}{c|}{3.40e--3} &
  6.29e--3 &
  \multicolumn{1}{c|}{\cellcolor[HTML]{C9C9C9}--4.17e--4} &
  \multicolumn{1}{c|}{2.37e--3} &
  4.33e--3 &
  \multicolumn{1}{c|}{\cellcolor[HTML]{C9C9C9}--5.81e--4} &
  \multicolumn{1}{c|}{\cellcolor[HTML]{C9C9C9}--4.06e--4} &
  6.83e--3 &
  \multicolumn{1}{c|}{\cellcolor[HTML]{C9C9C9}--3.68e--4} &
  \multicolumn{1}{c|}{\cellcolor[HTML]{C9C9C9}--1.86e--3} &
  \cellcolor[HTML]{C9C9C9}--1.94e--4 \\ \hline
TD &
  \multicolumn{1}{c|}{2.23e--5} &
  \multicolumn{1}{c|}{1.99e--3} &
  5.46e--3 &
  \multicolumn{1}{c|}{5.16e--5} &
  \multicolumn{1}{c|}{1.40e--3} &
  5.06e--3 &
  \multicolumn{1}{c|}{\cellcolor[HTML]{C9C9C9}--1.51e--4} &
  \multicolumn{1}{c|}{\cellcolor[HTML]{C9C9C9}--2.65e--3} &
  5.21e--3 &
  \multicolumn{1}{c|}{\cellcolor[HTML]{C9C9C9}--3.81e--4} &
  \multicolumn{1}{c|}{\cellcolor[HTML]{C9C9C9}--3.92e--4} &
  9.53e--3 \\ \hline
\end{tabular}
\label{tab:difference_controllability4}
\end{table*}

\begin{table*}[htbp]
\centering
\caption{Difference of MAE values of connectivity prediction between the null-mask and confusion-mask information loss, when $\langle k\rangle=7$.}
\begin{tabular}{|c|ccc|ccc|ccc|ccc|}
\hline
 &
  \multicolumn{3}{c|}{ER} &
  \multicolumn{3}{c|}{QS} &
  \multicolumn{3}{c|}{SW} &
  \multicolumn{3}{c|}{SF} \\ \cline{2-13} 
\multirow{-2}{*}{} &
  \multicolumn{1}{c|}{54} &
  \multicolumn{1}{c|}{216} &
  360 &
  \multicolumn{1}{c|}{54} &
  \multicolumn{1}{c|}{216} &
  360 &
  \multicolumn{1}{c|}{54} &
  \multicolumn{1}{c|}{216} &
  360 &
  \multicolumn{1}{c|}{54} &
  \multicolumn{1}{c|}{216} &
  360 \\ \hline
RA &
  \multicolumn{1}{c|}{\cellcolor[HTML]{C9C9C9}--1.97e--5} &
  \multicolumn{1}{c|}{4.21e--4} &
  3.85e--3 &
  \multicolumn{1}{c|}{\cellcolor[HTML]{C9C9C9}--3.84e--4} &
  \multicolumn{1}{c|}{2.13e--4} &
  6.77e--3 &
  \multicolumn{1}{c|}{\cellcolor[HTML]{C9C9C9}--5.19e--5} &
  \multicolumn{1}{c|}{9.03e--4} &
  5.82e--3 &
  \multicolumn{1}{c|}{1.26e--4} &
  \multicolumn{1}{c|}{2.08e--3} &
  7.36e--3 \\ \hline
TB &
  \multicolumn{1}{c|}{1.20e--3} &
  \multicolumn{1}{c|}{1.31e--3} &
  \cellcolor[HTML]{C9C9C9}--1.51e--2 &
  \multicolumn{1}{c|}{9.93e--5} &
  \multicolumn{1}{c|}{1.81e--3} &
  \cellcolor[HTML]{C9C9C9}--1.09e--3 &
  \multicolumn{1}{c|}{\cellcolor[HTML]{C9C9C9}--7.34e--4} &
  \multicolumn{1}{c|}{2.73e--3} &
  5.75e--3 &
  \multicolumn{1}{c|}{\cellcolor[HTML]{C9C9C9}--1.01e--5} &
  \multicolumn{1}{c|}{\cellcolor[HTML]{C9C9C9}--1.09e--2} &
  2.57e--3 \\ \hline
TD &
  \multicolumn{1}{c|}{\cellcolor[HTML]{C9C9C9}--1.86e--4} &
  \multicolumn{1}{c|}{1.67e--3} &
  1.18e--2 &
  \multicolumn{1}{c|}{\cellcolor[HTML]{C9C9C9}--1.48e--3} &
  \multicolumn{1}{c|}{1.08e--3} &
  1.19e--2 &
  \multicolumn{1}{c|}{\cellcolor[HTML]{C9C9C9}--7.85e--4} &
  \multicolumn{1}{c|}{1.75e--3} &
  9.01e--3 &
  \multicolumn{1}{c|}{8.63e--5} &
  \multicolumn{1}{c|}{\cellcolor[HTML]{C9C9C9}--1.36e--3} &
  1.19e--3 \\ \hline
\end{tabular}
\label{tab:difference_connectivity7}
\end{table*}

\begin{table*}[htbp]
\centering
\caption{Difference of MAE values of controllability prediction between the null-mask and confusion-mask information loss, when $\langle k\rangle=7$.}
\begin{tabular}{|c|ccc|ccc|ccc|ccc|}
\hline
 &
  \multicolumn{3}{c|}{ER} &
  \multicolumn{3}{c|}{QS} &
  \multicolumn{3}{c|}{SW} &
  \multicolumn{3}{c|}{SF} \\ \cline{2-13} 
\multirow{-2}{*}{} &
  \multicolumn{1}{c|}{54} &
  \multicolumn{1}{c|}{216} &
  360 &
  \multicolumn{1}{c|}{54} &
  \multicolumn{1}{c|}{216} &
  360 &
  \multicolumn{1}{c|}{54} &
  \multicolumn{1}{c|}{216} &
  360 &
  \multicolumn{1}{c|}{54} &
  \multicolumn{1}{c|}{216} &
  360 \\ \hline
RA &
  \multicolumn{1}{c|}{\cellcolor[HTML]{C9C9C9}--1.47e--4} &
  \multicolumn{1}{c|}{1.88e--2} &
  1.11e--2 &
  \multicolumn{1}{c|}{\cellcolor[HTML]{C9C9C9}--2.33e--4} &
  \multicolumn{1}{c|}{1.99e--2} &
  1.74e--2 &
  \multicolumn{1}{c|}{2.40e--5} &
  \multicolumn{1}{c|}{1.81e--2} &
  2.04e--3 &
  \multicolumn{1}{c|}{3.15e--4} &
  \multicolumn{1}{c|}{1.44e--2} &
  7.95e--3 \\ \hline
TB &
  \multicolumn{1}{c|}{1.42e--5} &
  \multicolumn{1}{c|}{3.54e--3} &
  5.53e--3 &
  \multicolumn{1}{c|}{\cellcolor[HTML]{C9C9C9}--7.60e--4} &
  \multicolumn{1}{c|}{3.50e--3} &
  4.60e--3 &
  \multicolumn{1}{c|}{1.99e--4} &
  \multicolumn{1}{c|}{2.37e--3} &
  8.68e--4 &
  \multicolumn{1}{c|}{\cellcolor[HTML]{C9C9C9}--9.95e--4} &
  \multicolumn{1}{c|}{8.92e--4} &
  9.56e--3 \\ \hline
TD &
  \multicolumn{1}{c|}{2.14e--4} &
  \multicolumn{1}{c|}{3.09e--3} &
  3.79e--3 &
  \multicolumn{1}{c|}{\cellcolor[HTML]{C9C9C9}--9.77e--5} &
  \multicolumn{1}{c|}{2.81e--3} &
  5.52e--3 &
  \multicolumn{1}{c|}{1.44e--4} &
  \multicolumn{1}{c|}{2.01e--3} &
  3.80e--3 &
  \multicolumn{1}{c|}{\cellcolor[HTML]{C9C9C9}--5.57e--4} &
  \multicolumn{1}{c|}{4.44e--4} &
  9.18e--3 \\ \hline
\end{tabular}
\label{tab:difference_controllability7}
\end{table*}

\begin{table*}[htbp]
\centering
\caption{Difference of MAE values of connectivity prediction between the null-mask and confusion-mask information loss, when $\langle k\rangle=9$.}
\begin{tabular}{|c|ccc|ccc|ccc|ccc|}
\hline
 &
  \multicolumn{3}{c|}{ER} &
  \multicolumn{3}{c|}{QS} &
  \multicolumn{3}{c|}{SW} &
  \multicolumn{3}{c|}{SF} \\ \cline{2-13} 
\multirow{-2}{*}{} &
  \multicolumn{1}{c|}{54} &
  \multicolumn{1}{c|}{216} &
  360 &
  \multicolumn{1}{c|}{54} &
  \multicolumn{1}{c|}{216} &
  360 &
  \multicolumn{1}{c|}{54} &
  \multicolumn{1}{c|}{216} &
  360 &
  \multicolumn{1}{c|}{54} &
  \multicolumn{1}{c|}{216} &
  360 \\ \hline
RA &
  \multicolumn{1}{c|}{\cellcolor[HTML]{C9C9C9}--2.25e--5} &
  \multicolumn{1}{c|}{3.91e--4} &
  5.20e--3 &
  \multicolumn{1}{c|}{\cellcolor[HTML]{C9C9C9}--2.38e--4} &
  \multicolumn{1}{c|}{2.49e--3} &
  6.24e--3 &
  \multicolumn{1}{c|}{7.29e--6} &
  \multicolumn{1}{c|}{8.77e--5} &
  5.30e--3 &
  \multicolumn{1}{c|}{\cellcolor[HTML]{C9C9C9}--6.20e--5} &
  \multicolumn{1}{c|}{1.07e--3} &
  1.02e--2 \\ \hline
TB &
  \multicolumn{1}{c|}{1.99e--3} &
  \multicolumn{1}{c|}{1.78e--3} &
  \cellcolor[HTML]{C9C9C9}--1.55e--3 &
  \multicolumn{1}{c|}{3.53e--4} &
  \multicolumn{1}{c|}{1.50e--4} &
  9.51e--3 &
  \multicolumn{1}{c|}{8.30e--4} &
  \multicolumn{1}{c|}{\cellcolor[HTML]{C9C9C9}--2.44e--3} &
  \cellcolor[HTML]{C9C9C9}--5.03e--3 &
  \multicolumn{1}{c|}{\cellcolor[HTML]{C9C9C9}--4.09e--4} &
  \multicolumn{1}{c|}{\cellcolor[HTML]{C9C9C9}--9.79e--4} &
  2.68e--2 \\ \hline
TD &
  \multicolumn{1}{c|}{8.51e--5} &
  \multicolumn{1}{c|}{6.72e--3} &
  1.35e-2 &
  \multicolumn{1}{c|}{\cellcolor[HTML]{C9C9C9}--1.21e--3} &
  \multicolumn{1}{c|}{6.30e--3} &
  9.74e--3 &
  \multicolumn{1}{c|}{\cellcolor[HTML]{C9C9C9}--8.92e--4} &
  \multicolumn{1}{c|}{7.96e--3} &
  7.77e--3 &
  \multicolumn{1}{c|}{1.87e--4} &
  \multicolumn{1}{c|}{7.20e--4} &
  6.80e--3 \\ \hline
\end{tabular}
\label{tab:difference_connectivity9}
\end{table*}

\begin{table*}[htbp]
\centering
\caption{Difference of MAE values of controllability prediction between the null-mask and confusion-mask information loss, when $\langle k\rangle=9$.}
\begin{tabular}{|c|ccc|ccc|ccc|ccc|}
\hline
 &
  \multicolumn{3}{c|}{ER} &
  \multicolumn{3}{c|}{QS} &
  \multicolumn{3}{c|}{SW} &
  \multicolumn{3}{c|}{SF} \\ \cline{2-13} 
\multirow{-2}{*}{} &
  \multicolumn{1}{c|}{54} &
  \multicolumn{1}{c|}{216} &
  360 &
  \multicolumn{1}{c|}{54} &
  \multicolumn{1}{c|}{216} &
  360 &
  \multicolumn{1}{c|}{54} &
  \multicolumn{1}{c|}{216} &
  360 &
  \multicolumn{1}{c|}{54} &
  \multicolumn{1}{c|}{216} &
  360 \\ \hline
RA &
  \multicolumn{1}{c|}{\cellcolor[HTML]{C9C9C9}--1.47e--4} &
  \multicolumn{1}{c|}{1.90e--2} &
  \cellcolor[HTML]{C9C9C9}--1.57e--3 &
  \multicolumn{1}{c|}{\cellcolor[HTML]{C9C9C9}--5.43e--4} &
  \multicolumn{1}{c|}{1.82e--2} &
  2.77e--3 &
  \multicolumn{1}{c|}{\cellcolor[HTML]{C9C9C9}--3.40e--4} &
  \multicolumn{1}{c|}{1.60e--2} &
  \cellcolor[HTML]{C9C9C9}--1.68e--3 &
  \multicolumn{1}{c|}{3.67e--4} &
  \multicolumn{1}{c|}{1.06e--2} &
  3.88e--4 \\ \hline
TB &
  \multicolumn{1}{c|}{8.50e--6} &
  \multicolumn{1}{c|}{4.71e--3} &
  \cellcolor[HTML]{C9C9C9}--6.51e--3 &
  \multicolumn{1}{c|}{3.55e--4} &
  \multicolumn{1}{c|}{4.44e--3} &
  \cellcolor[HTML]{C9C9C9}--3.64e--3 &
  \multicolumn{1}{c|}{4.79e--4} &
  \multicolumn{1}{c|}{4.26e--3} &
  \cellcolor[HTML]{C9C9C9}--7.22e--3 &
  \multicolumn{1}{c|}{\cellcolor[HTML]{C9C9C9}--7.97e--4} &
  \multicolumn{1}{c|}{\cellcolor[HTML]{C9C9C9}--2.02e--3} &
  8.77e--3 \\ \hline
TD &
  \multicolumn{1}{c|}{1.74e--4} &
  \multicolumn{1}{c|}{3.53e--3} &
  \cellcolor[HTML]{C9C9C9}--1.69e--5 &
  \multicolumn{1}{c|}{9.75e--5} &
  \multicolumn{1}{c|}{3.12e--3} &
  1.28e--3 &
  \multicolumn{1}{c|}{1.38e--4} &
  \multicolumn{1}{c|}{3.29e--3} &
  2.53e--3 &
  \multicolumn{1}{c|}{\cellcolor[HTML]{C9C9C9}--4.88e--4} &
  \multicolumn{1}{c|}{\cellcolor[HTML]{C9C9C9}--3.09e--3} &
  6.83e--3 \\ \hline
\end{tabular}
\label{tab:difference_controllability9}
\end{table*}

\begin{table}[htbp]
\centering
\caption{Quantitative statistics of the positive and negative MAE difference values and the average values.}
\begin{tabular}{|c|cc|cc|}
\hline & \multicolumn{2}{c|}{connectivity} & \multicolumn{2}{c|}{controllability} \\ \cline{2-5} 
\multirow{-2}{*}{} & \multicolumn{1}{c|}{+} & -- & \multicolumn{1}{c|}{+} & -- \\ \hline
$\langle k\rangle$=4 &
  \multicolumn{1}{c|}{\cellcolor[HTML]{C9C9C9}20 (3.48e--3)} & 16 (--5.87e--3) &
  \multicolumn{1}{c|}{\cellcolor[HTML]{C9C9C9}21 (8.38e--3)} & 15 (--1.12e--3) \\ \hline
$\langle k\rangle$=7 &
  \multicolumn{1}{c|}{\cellcolor[HTML]{C9C9C9}24 (3.39e--3)} & 12 (--2.67e--3) &
  \multicolumn{1}{c|}{\cellcolor[HTML]{C9C9C9}30 (5.73e--3)} & 6 (--4.65e--4) \\ \hline
$\langle k\rangle$=9 &
  \multicolumn{1}{c|}{\cellcolor[HTML]{C9C9C9}26 (5.08e--3)} & 10 (--1.28e--3) &
  \multicolumn{1}{c|}{\cellcolor[HTML]{C9C9C9}23 (4.84e--3)} & 13 (--2.15e--3) \\ \hline
\end{tabular}
\label{tab:statistic_error}
\end{table}

\begin{figure*}[htbp]
	\centering
	\includegraphics[width=0.8\linewidth]{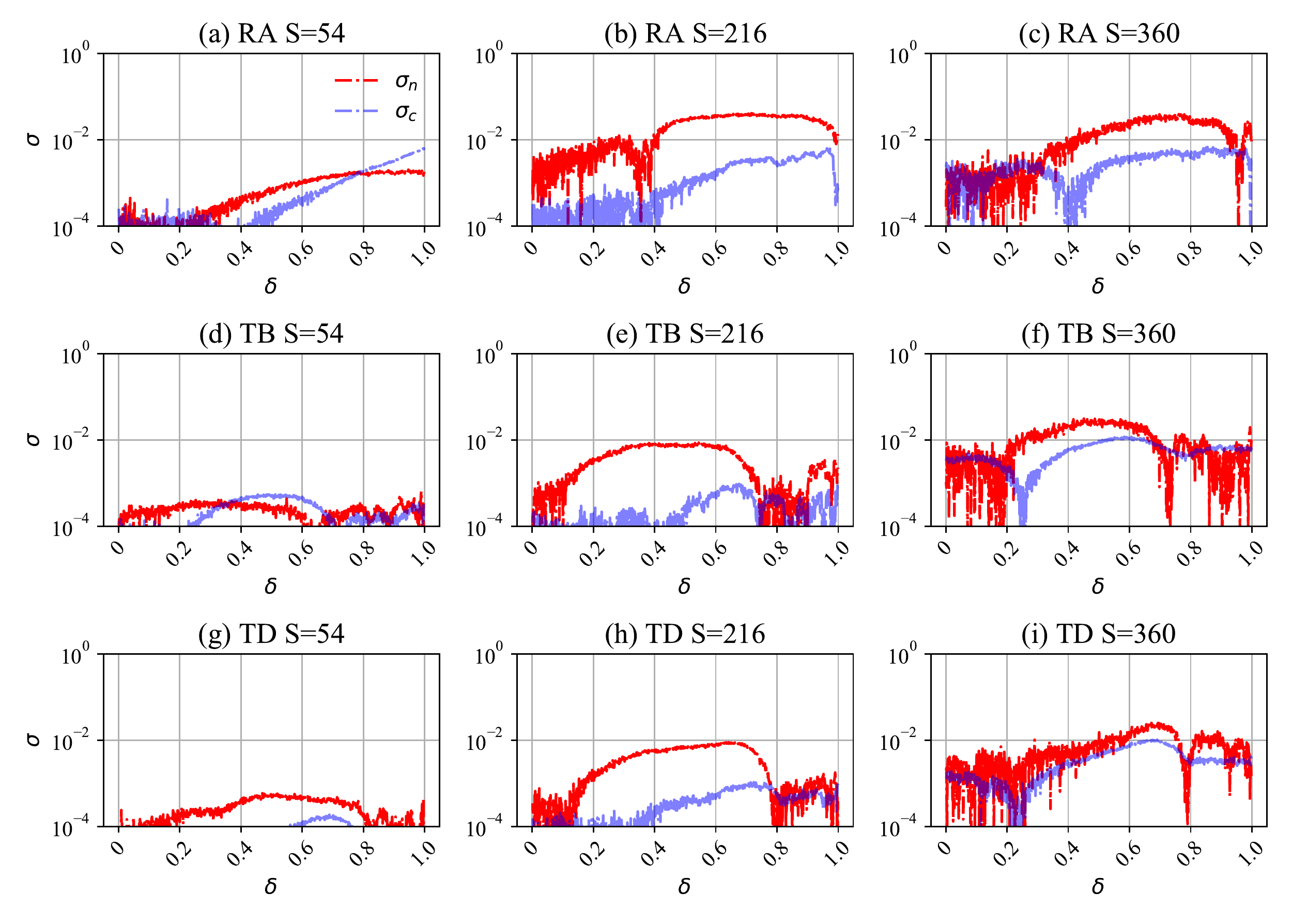}
	\caption{The absolute prediction errors when null masks and confusion masks are used. $\delta$ represents the proportion of removed nodes; $\sigma_n$ represents the absolute difference between the prediction errors with and without null masks; $\sigma_c$ means that for confusion masks; $S$ represents the mask size. ER networks with $\langle k\rangle=7$ are adopted. }
	\label{fig:controllability_er_7_compare}
\end{figure*}

Tables \ref{tab:difference_connectivity4}--\ref{tab:difference_controllability9} show the difference of MAE values between the prediction errors when null masks and confusion masks are used for information loss.  
In these tables, a positive value means that the prediction error obtained when there is null masks is lower than the prediction error obtained when there is a confusion mask; while a negative value means a higher prediction error is obtained when a null mask is used. 
To present the data clearly, negative values in Tables \ref{tab:difference_connectivity4}--\ref{tab:difference_controllability9} are highlighted in gray.  Overall, there are more positive values (144 positive values and 72 negative values in total) in these tables, meaning that using confusion masks help CNN-RP to obtain lower prediction error, compared to using null masks. 
This means that when confusion masks are used, CNN-RP is more tolerable to the network information loss.

Fig. \ref{fig:controllability_er_7_compare} shows the comparison of absolute prediction errors obtained by CNN-RP, when null masks and confusion masks are used.  It is clear that red curves are mostly higher then blue curves, meaning that when null masks are used, the prediction errors are generally higher.  This observation is consistent with that from Tables \ref{tab:difference_connectivity4}--\ref{tab:difference_controllability9}.

The overall prediction error performance is summarized in Table \ref{tab:statistic_error}. 
The positive and negative values in Tables \ref{tab:difference_connectivity4}--\ref{tab:difference_controllability9} are summed up and the numbers of positive and negative values are counted. Again, using confusion masks to represent information loss is less misleading for CNN-RP than using null masks, for both connectivity and controllability robustness predictions.

\subsection{Discussions} 
In Experiment I, different amounts of missing edges or information loss are simulated by different sizes of mask. When the mask size is small, the prediction performance of CNN-RP is slightly affected, while if the mask size exceeds a threshold, then its influence to prediction performance becomes significant. In this case, when null masks are used, CNN-RP can not determine the missing edges and treats the mask-covered area as a part of the true topology. 

In contrast, when confusion masks are applied, CNN-RP is possible to distinguish the masked area. In Experiment II, both the original and masked networks are employed in the training, which enhances the ability of CNN-RP to distinguish the masked area from the `true network topology'.  By using a mixed training set where both confusion-masked and original networks are employed, CNN-RP is able to treat (or ignore) the masked area properly, which is reflected by obtaining lower prediction errors.

\section{Conclusions}\label{sec:end}

Connectivity and controllability robustness are important indicators of complex networks.  They can be indicated using a sequence of connectivity and controllability measures during node-removal attacks as shown in Equations (\ref{eq:nlc}) and (\ref{eq:ndi}). CNN-RP has good performance for network robustness prediction when the complete information of a network is known. However, missing information such as missing edges is inevitable in real-world network applications. The missing information may have a significant impact to revealing network functions such as robustness.
To investigate the effect of missing edges to CNN-RP prediction performance on network robustness, masks are implemented to simulate missing edges, by covering edge information within a masked square area in the network adjacency matrix.  Two mask types including null masks and confusion masks are compared. Missing edges are marked `no edges' in null masks and marked as `unknown' in confusion masks. Different null mask sizes are also implemented and compared. 
A threshold is explored that if a total amount of 7.29\% (or more) information is lost, the performance of CNN-based prediction will be significantly degenerated. 
Experimental results also reveal that null masks are misleading to the CNN-based predictors. CNN-RP is able to learn some features of missing edges and well predict the robustness performance, if the missing edge information can be denoted by special markers, under the confusion mask strategy.

\bibliographystyle{IEEEtran}
\bibliography{ref}

\end{document}